\documentclass[aps,prd,twocolumn,showpacs,amsmath,amssymb]{revtex4-1}
\usepackage{amsmath}
\usepackage{graphicx}
\usepackage{subfigure}
\usepackage{epstopdf}
\usepackage{color}
\usepackage{multirow}
\usepackage{setspace}
\usepackage{overpic}
\usepackage{amssymb}
\usepackage[bookmarksnumbered, pdfstartview=FitH,colorlinks,urlcolor=blue, citecolor=blue,linkcolor=blue] {hyperref}
\usepackage{lineno}
\usepackage{bm}
\usepackage{rotating}
\usepackage[utf8]{inputenc}

\usepackage{tikz}

\let\oldequation\equation
\let\oldendequation\endequation

\renewenvironment{equation}
  {\linenomathNonumbers\oldequation}
  {\oldendequation\endlinenomath}

\begin{document}

\title{\bf \boldmath
Improved measurement of the semileptonic decay $D^+_{s}\to K^0 e^+\nu_e$
}

\author{M.~Ablikim$^{1}$, M.~N.~Achasov$^{4,c}$, P.~Adlarson$^{76}$, O.~Afedulidis$^{3}$, X.~C.~Ai$^{81}$, R.~Aliberti$^{35}$, A.~Amoroso$^{75A,75C}$, Q.~An$^{72,58,a}$, Y.~Bai$^{57}$, O.~Bakina$^{36}$
% ...
\\
\vspace{0.2cm}
(BESIII Collaboration)\\
\vspace{0.2cm} {\it
$^{1}$ Institute of High Energy Physics, Beijing 100049, People's Republic of China\\
% ...
$^{3}$ Bochum Ruhr-University, D-44780 Bochum, Germany\\
$^{4}$ Budker Institute of Nuclear Physics SB RAS (BINP), Novosibirsk 630090, Russia\\
% ...
$^{35}$ Johannes Gutenberg University of Mainz, Johann-Joachim-Becher-Weg 45, D-55099 Mainz, Germany\\
$^{36}$ Joint Institute for Nuclear Research, 141980 Dubna, Moscow region, Russia\\
% ...
$^{57}$ Southeast University, Nanjing 211100, People's Republic of China\\
$^{58}$ State Key Laboratory of Particle Detection and Electronics, Beijing 100049, Hefei 230026, People's Republic of China\\
% ...
$^{72}$ University of Science and Technology of China, Hefei 230026, People's Republic of China\\
% ...
$^{75}$ University of Turin and INFN, (A)University of Turin, I-10125, Turin, Italy; (B)University of Eastern Piedmont, I-15121, Alessandria, Italy; (C)INFN, I-10125, Turin, Italy\\
$^{76}$ Uppsala University, Box 516, SE-75120 Uppsala, Sweden\\
% ...
$^{81}$ Zhengzhou University, Zhengzhou 450001, People's Republic of China\\
\vspace{0.2cm}
$^{a}$ Deceased\\
% ...
$^{c}$ Also at the Novosibirsk State University, Novosibirsk, 630090, Russia\\
% ...
}
}

\begin{abstract}
Analyzing $e^+e^-$ collision data corresponding to an integrated luminosity of $7.33~\mathrm{fb}^{-1}$ collected at center-of-mass energies between 4.128 and 4.226~GeV with the BESIII detector, we measure the branching fraction of the semileptonic decay $D^+_{s}\to K^0 e^+\nu_e$ to be $(2.98\pm0.23\pm0.12)\times10^{-3}$. The $D_s^+\to K^0$ hadronic form factor is determined from the differential decay rate of $D^+_s\to K^0 e^+\nu_e$ to be $f^{K^0}_+(0)=0.636\pm0.049\pm0.013$. For both measurements, the first uncertainty is statistical and the second systematic. The branching fraction and form factor measurements are factors of 1.6 and 1.7 more precise than the previous world averages, respectively.
\end{abstract}

\maketitle
\section{Introduction}
Studies of semileptonic $D^+_s$ decays provide important input to understand the effects of the strong and weak interactions in charmed meson decays~\cite{Li:2021iwf}.
The partial decay rates of the semileptonic decays $D^+_s\to P\ell^+\nu_\ell$ ($P$ denotes a pseudoscalar meson)
are  proportional  to the product of the hadronic form factor $f^{P}_{+}(0)$ and the Cabibbo-Kobayashi-Maskawa~(CKM) matrix element $|V_{cs}|$ or $|V_{cd}|$.
In recent years, there has been much progress in the experimental study of semileptonic $D^+_s$ decays. However, knowledge of Cabibbo-suppressed semileptonic $D_{s}^{+}$ decays remains limited by statistical uncertainty~\cite{PDG2022}.
Improved measurements of the branching fraction of $D^+_s\to K^0 e^+\nu_e$ and the hadronic form factor of $D^+_s\to K^0$ are important to validate theoretical calculations~\cite{Wuyl,ly,cheng,Melikhov,Vermayw,soni1,soni2,Fajfer,rqm}.
The hadronic form factor measurement helps test and improve theoretical calculations, which in turn improves the measured precision of $|V_{cd}|$. This is important for testing the unitary of the CKM matrix and searching for possible indications of new physics.

Theoretical predictions of the branching fraction of $D^+_s\to K^0 e^+\nu_e$ range from $2.0\times10^{-3}$ to $4.0\times10^{-3}$. In 2009, the CLEO-c experiment reported the first measurement of the branching fraction of $D^+_s\to K^0 e^+\nu_e$ using 0.31 fb$^{-1}$ of $e^+e^-$ collision data collected at a center-of-mass (CM) energy of 4.17 GeV~\cite{cleo-K0enu}. In 2015, the CLEO collaboration updated the branching fraction measurement using 0.586 fb$^{-1}$ of data at the same energy point~\cite{Hietala}.
In 2019, the BESIII experiment presented a further improved measurement of the branching fraction and the first measurement of the hadronic form factor in $D^+_s\to K^0 e^+\nu_e$ by analyzing 3.19~fb$^{-1}$ of data at 4.178 GeV~\cite{bes3-K0enu,Ke:2023qzc}.

In this paper, we report improved measurements of both the branching fraction and the hadronic transition form factor in $D_s^+\to K^0e^+\nu_e$, assuming $K^0 \to K_S^0$ with a branching fraction of 50\%, based on $7.33~\mathrm{fb}^{-1}$ of $e^+e^-$ collision data taken with center-of-mass (CM) energies between 4.128 and 4.226 GeV with the BESIII detector.
Throughout this paper, charge conjugate modes are implied.

\section{BESIII detector and Monte Carlo simulations}

The BESIII detector~\cite{Ablikim:2009aa} records symmetric $e^+e^-$ collisions provided by the BEPCII storage ring~\cite{Yu:IPAC2016-TUYA01}, which operates with a peak luminosity of $1.1\times10^{33}$~cm$^{-2}$s$^{-1}$ in the CM energy range from 1.84 to 4.95~GeV.
BESIII has collected large data samples in this energy region~\cite{Ablikim:2019hff}. The cylindrical core of the BESIII detector covers 93\% of the full solid angle and consists of a helium-based multilayer drift chamber~(MDC), a plastic scintillator time-of-flight system~(TOF), and a CsI(Tl) electromagnetic calorimeter~(EMC), all enclosed in a superconducting solenoidal magnet that provides a 1.0~T magnetic field.
The solenoid is supported by an octagonal flux-return yoke with resistive plate counter muon
identification modules interleaved with steel.
The charged-particle momentum resolution at $1~{\rm GeV}/c$ is
$0.5\%$, and the d$E$/d$x$ resolution is $6\%$ for electrons
from Bhabha scattering. The EMC measures photon energies with a
resolution of $2.5\%$ ($5\%$) at $1$~GeV in the barrel (end cap) region. The time resolution in the TOF barrel (end cap) region is 68 (110)~ps, and the end cap TOF system was upgraded in 2015 using multi-gap resistive plate chamber technology, providing a time resolution of 60~ps~\cite{etof}. Approximately 83\% of the data benefits from this upgrade.

Simulated data samples produced with a {\sc
geant4}-based~\cite{geant4} Monte Carlo (MC) package, which
includes the geometric description of the BESIII detector and the detector response, are used to determine detection efficiencies and to estimate backgrounds. The simulation models the beam energy spread and initial state radiation (ISR) in the $e^+e^-$ annihilations with the generator {\sc
kkmc}~\cite{ref:kkmc}.
In the simulation, the production of open-charm
processes directly produced via $e^+e^-$ annihilations are modeled with the generator {\sc conexc}~\cite{ref:conexc},
and their subsequent decays are modeled by {\sc evtgen}~\cite{ref:evtgen} with known branching fractions from the Particle Data Group~(PDG)~\cite{PDG2022}.
The ISR production of vector charmonium(-like) states
and the continuum processes are incorporated in {\sc
kkmc}~\cite{ref:kkmc}.
The remaining unknown charmonium decays
are modeled with {\sc lundcharm}~\cite{ref:lundcharm}. Final state radiation (FSR) from charged final-state particles is incorporated using the {\sc photos} package~\cite{photos}.

\section{Analysis method}

Pairs of $D_s^{*\pm}D_s^\mp$ decaying into $\gamma D_s^+ D_s^-$ are produced copiously in $e^+e^-$ collisions with CM energies between 4.128 and 4.226~GeV.
This allows us to study $D_s^+$ decays using the double-tag (DT) method pioneered by the MARK-III collaboration~\cite{DTmethod}.
The $D_s^-$ meson, which is fully reconstructed via one of its hadronic decay modes, is referred to as a single-tag (ST) $D_s^-$ meson.
In the presence of a fully reconstructed ST $D_s^-$ meson at a certain CM energy, we can infer the kinematic information of the other $D_s^+$ meson. The semileptonic decay $D_s^+ \to K^0 e^+ \nu$ is thus selected on the side recoiling against the ST $D_s^-$, despite the presence of an undetectable neutrino.
A DT event is an event in which the transition $\gamma$ from the $D_s^{*+}$ and the semileptonic decay $D_s^+\to K^0 e^+\nu_e$ can be successfully selected in the presence of the ST.
The branching fraction of $D^+_s\to K^0e^+\nu_e$ is determined by
\begin{equation}
\mathcal B_{D_s^+\to K^0e^+\nu_e}=\frac{N_{\rm DT}}{N_{\rm ST}^{\rm tot} \cdot \bar \epsilon_{\gamma\rm SL}},
\label{eq1}
\end{equation}
where $N_{\rm DT} = \sum_{ij}N_{\rm DT}^{ij}$ and $N_{\rm ST}^{\rm tot} = \sum_{ij}N_{\rm ST}^{ij}$ are the yields of the DT events and ST $D^-_s$ mesons in data summing over all tag modes $i$ and datasets $j$, respectively; and $\bar \epsilon_{\gamma\rm SL}$ is the efficiency of detecting the $\gamma$ and the semileptonic decay in the presence of the ST $D^-_s$ candidate, weighted by the ST yield in data. It is calculated by $\sum_{ij}[(N_{\rm}^{ij}/N_{\rm ST})\cdot(\epsilon_{\rm DT}^{ij}/\epsilon_{\rm ST}^{ij})]$, where $\epsilon_{\rm DT}^{ij}$ and $\epsilon_{\rm ST}^{ij}$ are the detection efficiencies of the DT and ST candidates, respectively.

\section{Single-tag $D^-_s$ candidates}

The ST $D^-_s$ candidates are formed using fourteen hadronic decay modes:
$D^-_s\to K^+K^-\pi^-$, $K^+K^-\pi^-\pi^0$, $\pi^+\pi^-\pi^-$, $K^0_SK^-$, $K^0_SK^-\pi^0$,$K^-\pi^+\pi^-$, $K^0_SK^0_S\pi^-$,\,$K^0_SK^+\pi^-\pi^-$, $K^0_SK^-\pi^+\pi^-$, $\eta_{\gamma\gamma}\pi^-$, $\eta^\prime_{\pi^+\pi^-\eta}\rho^-$, $\eta^\prime_{\eta_{\gamma\gamma}\pi^+\pi^-}\pi^-$, $\eta^\prime_{\gamma\rho^0}\pi^-$, and $\eta_{\gamma\gamma}\rho^-$.
Throughout this paper,
the subscripts on the $\eta^{(\prime)}$ denote the decay modes that are used to reconstruct the $\eta^{(\prime)}$ candidates
and $\rho$ denotes $\rho(770)$.

In selecting candidates for the $K^\pm$, $\pi^\pm$, $K^0_S$, $\gamma$, $\pi^0$, and $\eta$, we use the same selection criteria as those adopted in our previous works~\cite{bes3_etaev,bes3_gev}.
All charged tracks, except for those from $K^0_S$ decays, are required to originate from a region defined as
$|V_{xy}|<1$ cm, $|V_{z}|<10$ cm, and $|\!\cos\theta|<0.93$,
where $|V_{xy}|$ and $|V_{z}|$ are the distances of closest approach to the interaction point~(IP) in the transverse plane and
along the MDC axis, respectively, and $\theta$ is the polar angle with respect to the MDC axis.
The particle identification (PID) of charged particles is performed with combined $dE/dx$ and TOF information.
Those with confidence level for the pion (kaon) hypothesis greater than that for the kaon (pion) hypothesis are assigned
to be pion (kaon) candidates.

Candidates for $K_{S}^0$ are reconstructed from two oppositely charged tracks satisfying $|V_{z}|<$ 20~cm. The two charged tracks are assigned as $\pi^+\pi^-$ without imposing further PID criteria. They are constrained to
originate from a common vertex and are required to have an invariant mass
within $|M_{\pi^{+}\pi^{-}} - m_{K_{S}^{0}}|<$ 12~MeV$/c^{2}$, where
$m_{K_{S}^{0}}$ is the $K^0_{S}$ nominal mass~\cite{PDG2022}. The
decay length of the $K^0_S$ candidate is required to be greater than
twice the vertex resolution away from the IP.

Photon candidates are selected using information measured by the EMC and are required to satisfy the following criteria.
To suppress backgrounds from electronic noise or bremsstrahlung,
any candidate shower is required to start within $[0, 700]$ ns from the event start time.
The energy of each shower in the barrel (endcap) region of the EMC~\cite{Ablikim:2009aa} is required to be greater than 25 (50) MeV.
To suppress backgrounds associated with charged tracks,
the minimum opening angle between the momentum of the candidate shower and the extrapolated momentum direction of the nearest charged track at the EMC has to be greater than $10^\circ$.

Candidates for $\pi^0$ and $\eta_{\gamma\gamma}$ are formed from $\gamma\gamma$ pairs with invariant masses in the mass intervals $(0.115,\,0.150)$ and $(0.50,\,0.57)$\,GeV$/c^{2}$, respectively.
To improve momentum resolution, the invariant mass of each selected $\gamma\gamma$ pair is constrained to either the $\pi^0$  or $\eta$ nominal mass~\cite{PDG2022}.
To form candidates for $\rho^{0(+)}$, $\eta_{\pi^0\pi^+\pi^-}$, $\eta^\prime_{\eta\pi^+\pi^-}$, and $\eta^\prime_{\gamma\rho^0}$,
the invariant masses of the $\pi^+\pi^{-(0)}$, $\pi^0\pi^+\pi^-$, $\eta\pi^+\pi^-$, and  $\gamma\rho^0$ combinations are required to be
within the mass intervals of $(0.57,\,0.97)$, $(0.53,\,0.57)$,  $(0.946,\,0.970)$, and $(0.940,\,0.976)~\mathrm{GeV}/c^2$, respectively.
In addition, the energy of the $\gamma$ resulting from the $\eta^\prime_{\gamma\rho^0}$ decay is required to be greater than 0.1\,GeV.

The background caused by the transition pions from $D^{*+}$ decays is suppressed by requiring the momentum of any pion which is not from a $K_S^0$, $\eta$, or $\eta^\prime$ to be greater than 0.1\,GeV/$c$.

The backgrounds from non-$D_s^{\pm}D^{*\mp}_s$ processes are suppressed with the beam-constrained mass of the ST $D_s^-$ candidate, which is defined as
\begin{equation}
M_{\rm BC}\equiv\sqrt{E^2_{\rm beam}/c^4-|\vec{p}_{\rm tag}|^2/c^2},
\end{equation}
where $E_{\rm beam}$ is the beam energy and
$\vec{p}_{\rm tag}$ is the momentum of the ST $D_s^-$ candidate in the $e^+e^-$ CM frame.
The $M_{\rm BC}$ is required to be within the intervals shown in Table~\ref{tab:mbc} .
This requirement retains 90\% of the $D_s^-$ and $D_s^+$ mesons from $e^+ e^- \to D_s^{*\mp}D_s^{\pm}$.

\begin{table}[htbp]
	\centering\linespread{1.15}
	\caption{Requirements on $M_{\rm BC}$ for various energy points.}
	\small
	\label{tab:mbc}
	\begin{tabular}{lc}
		\hline\hline
		$E_{\rm CM}$ (GeV) & $M_{\rm BC}$ (GeV/$c^2$) \\
		\hline
		4.128          & $[2.010,2.061]$         \\
		4.157          & $[2.010,2.070]$         \\
		4.178          & $[2.010,2.073]$         \\
		4.189          & $[2.010,2.076]$         \\
		4.199          & $[2.010,2.079]$         \\
		4.209          & $[2.010,2.082]$         \\
		4.219          & $[2.010,2.085]$         \\
		4.226          & $[2.010,2.088]$         \\
		\hline\hline
	\end{tabular}
\end{table}

\begin{figure*}[htbp]
	\centering
	\includegraphics[width=0.8\textwidth]{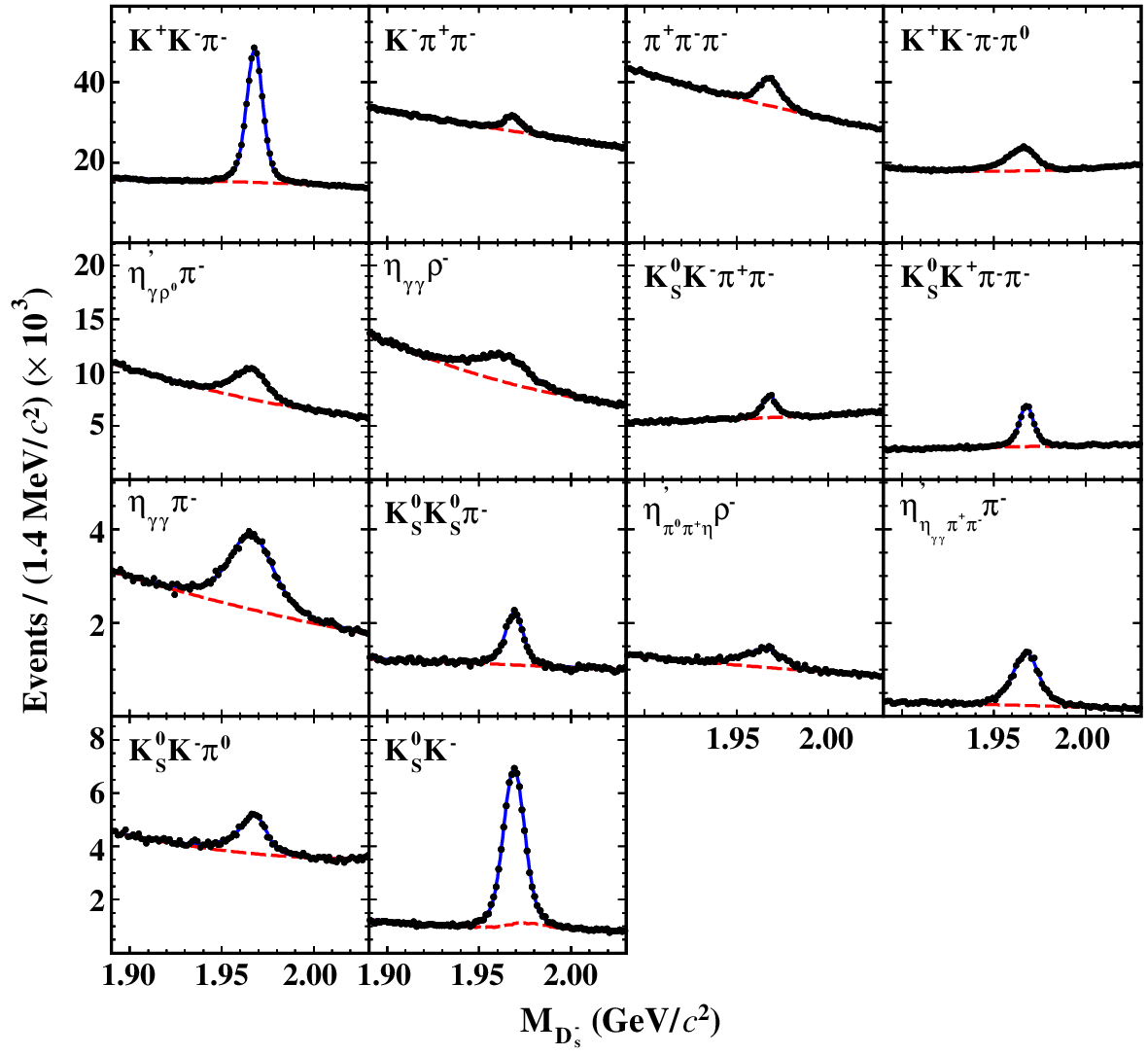}
	\caption{\footnotesize Fits to the $M_{D_s^-}$ distributions of the ST candidates for various tag modes.
		The points with error bars are data;
		the blue solid curves are the best fit results;
		and the red dashed curves are the fitted background shapes.
		}
	\label{fig:stfit}
\end{figure*}

In the case of multiple candidates, only the candidate with the $D_s^-$ recoil mass
\begin{equation}
M_{\rm rec} \equiv \sqrt{ \left (E_{\rm CM}/c^2 - \sqrt{|\vec p_{\rm tag}|^2/c^2+m^2_{D^-_s}} \right )^2
-|\vec p_{\rm tag}|^2/c^2}
\end{equation}
closest to the $D_s^{*+}$ nominal mass~\cite{PDG2022} per tag mode and per charge is retained for further analysis.
The distributions of the invariant mass ($M_{\rm tag}$) of the accepted ST candidates for various tag modes are shown in Fig.~\ref{fig:stfit}.
The yields of ST $D^-_s$ mesons reconstructed in various tag modes are derived from fits to individual $M_{\rm tag}$ distributions and are listed in Table~\ref{tab:bf}.
In the fits, the signal is described by the simulated signal shape convolved with a Gaussian function that represents the resolution difference between data and simulation.
In the fit to the $D^-_s\to K_S^0K^-$ tag mode,
the shape of the peaking background from $D^-\to K^0_S\pi^-$ is modeled by the simulated shape convolved with the same Gaussian resolution function as the signal shape and the size of this peaking background is free.
The combinatorial background is described by a second-order Chebychev polynomial function, and has been verified by analyzing the inclusive simulation sample.
Figure~\ref{fig:stfit} shows the results of the fit to the data sample.
For each tag mode, the ST yield is obtained by integrating the signal shape over the selected $D_s^-$ signal region defined within $1.94<M_{D_s^-}<1.99$~GeV/$c^2$.
The second and third columns of Table~\ref{tab:bf} summarize the yields of ST $D^-_s$ mesons ($N_{\rm ST}$) for various tag modes obtained from the data sample and the corresponding detection efficiencies ($\epsilon_{\rm ST}$), respectively. The total ST yield summed over all ST modes is $N_{\mathrm{ST}}^{\rm tot}=(783.1\pm2.5)\times 10^{-3}$, where the uncertainty is statistical only.

\begin{table}[htbp]
	\centering\linespread{1.15}
	\caption{
		Fitted yields of single-tag $D^-_s$ mesons from the data sample ($N_{\rm ST}$),
		the efficiencies of detecting single-tag $D^-_s$ mesons and double-tag events ($\epsilon_{\rm ST}$ and $\epsilon_{\rm DT}$) for various tag modes.
		For all quantities, the uncertainties are statistical only.
		The listed efficiencies do not include the branching fractions of the daughter particles's decays.}
	\small
	\label{tab:bf}
	\begin{tabular}{lr@{}lr@{}lr@{}l}\hline\hline
		Tag mode  &\multicolumn{2}{c}{$N_{\rm ST}$~($\times 10^3$)}&\multicolumn{2}{c}{$\epsilon_{\rm ST}$~(\%)}&\multicolumn{2}{c}{$\epsilon_{\rm DT}~(\%)$} \\ \hline
		$K^+K^-\pi^-$                      & 281.7   & $\pm$0.8   &41.94 & $\pm$0.03  &10.92 & $\pm$0.11\\
		$K^+K^-\pi^-\pi^0$                 & 85.4    & $\pm$1.0   &11.53 & $\pm$0.03  &3.39  & $\pm$0.12\\
		$\pi^-\pi^+\pi^-$                  & 76.9    & $\pm$1.0   &53.84 & $\pm$0.06  &15.28 & $\pm$0.11\\
		$K_S^0 K^-$                        & 63.2    & $\pm$0.3   &47.18 & $\pm$0.06  &12.73 & $\pm$0.11\\
		$K_S^0 K^-\pi^0$                   & 21.9    & $\pm$0.5   &16.56 & $\pm$0.08  &4.81  & $\pm$0.12\\
		$K^-\pi^+\pi^-$                    & 36.8    & $\pm$0.8   &47.06 & $\pm$0.08  &12.78 & $\pm$0.11\\
		$K_S^0 K_S^0 \pi^-$                & 10.6    & $\pm$0.2   &22.92 & $\pm$0.13  &5.82  & $\pm$0.12\\
		$K_S^0 K^+\pi^-\pi^-$              & 30.5    & $\pm$0.3   &21.48 & $\pm$0.07  &5.48  & $\pm$0.12\\
		$K_S^0 K^-\pi^+\pi^-$              & 16.4    & $\pm$0.4   &19.21 & $\pm$0.10  &4.74  & $\pm$0.12\\
		$\eta\pi^-$                        & 33.7    & $\pm$0.6   &40.43 & $\pm$0.08  &12.54 & $\pm$0.11\\
		$\eta'(\pi^+\pi^-\eta)\rho^-$      & 7.2     & $\pm$0.3   &5.62  & $\pm$0.08  &1.83  & $\pm$0.11\\
		$\eta'(\eta\pi^+\pi^-)\pi^-$       & 15.6    & $\pm$0.2   &19.20 & $\pm$0.10  &5.45  & $\pm$0.12\\
		$\eta'(\gamma\rho)\pi^-$           & 47.3    & $\pm$0.7   &30.94 & $\pm$0.07  &8.77  & $\pm$0.11\\
		$\eta\rho^-$                       & 56.0    & $\pm$1.2   &14.35 & $\pm$0.04  &5.02  & $\pm$0.12\\
		\hline\hline
	\end{tabular}
\end{table}

\section{Selection of $D_s^+\to K^0e^+\nu_e$}

In the system recoiling against the $D_s^-$ tag and the transition $\gamma$ from the $D_s^{*-}$,
the semileptonic decay $D_s^+\to K^0e^+\nu_e$ is selected using tracks that have not been used for the single tag reconstruction.
To identify positrons, the combined confidence levels $CL_{e}^\prime$, $CL_{\pi}^\prime$, and $CL_{K}^\prime$ for
the electron, pion, and kaon hypotheses are calculated with the ${\rm d}E/{\rm d}x$, TOF, and EMC information. The positron candidates are required to satisfy $CL_{e}^\prime >0.001$ and $CL_{e}^\prime / (CL_{e}^\prime + CL_{\pi}^\prime + CL_{K}^\prime ) >0.8$.

The background from $D_s^+\to K^0 K^+$ is vetoed by requiring the $K^0e^+$ invariant mass to be less than 1.78~GeV/$c^2$.
The background contributions from $D_s^+$ hadronic decays associated with fake photons misidentified from showers are rejected by requiring the largest energy of the unused showers($E_{\rm{extra} \gamma }^{\rm{max}}$) to be less than 0.2~GeV.

To identify the transition $\gamma$ produced directly from the $D_s^{*\pm}$, we perform kinematic fits under two hypotheses.
One assumes that the $D_s^{*-}$ is formed by the transition $\gamma$ and the ST $D^-_s$,
and the other assumes that the $D_s^{*+}$ is formed by the transition $\gamma$ and the semileptonic decay.
The final particles from the $D^{\mp}_sD_s^{*\pm}$ system are constrained to obey energy and momentum conservation in the $e^+e^-$ CM frame with the neutrino treated as a missing particle. The particle candidates for $D^{\pm}_s$ are constrained to their known mass from the PDG~\cite{PDG2022}.
For the former hypothesis, the mass of the transition $\gamma$ and the tagged $D^-_s$ is constrained to
the known $D_s^{*-}$ mass.
For the latter hypothesis, the mass of the transition $\gamma$ and the semileptonic decay is constrained to
the known $D_s^{*+}$ mass.
The hypothesis with the smallest $\chi^2$ of the kinematic fit($\chi^2_{\rm KMFIT}$), which also satisfies $\chi^2_{\rm KMFIT}<100$, is kept for further analysis.

The presence of the neutrino is inferred from the distribution of the missing-mass squared variable, which is defined as
\begin{equation}
M^{2}_{\rm miss}=E^{2}_{\rm miss}/c^4-|\vec p_{\rm miss}|^2/c^2.
\end{equation}
Here,
$E_{\rm miss}=E_{\rm CM}-\Sigma_i E_{i}$ and
$\vec p_{\rm miss}=\Sigma_i \vec{p}_{i}$,
where $E_{i}$ and $\vec{p}_{i}$, with $i =$ (tag, $\gamma, e$, and $K^0$), are the energy and momentum of particle $i$.

\begin{figure}[htbp]
	\centering
    \begin{minipage}[t]{7cm}
	\includegraphics[width=\textwidth]{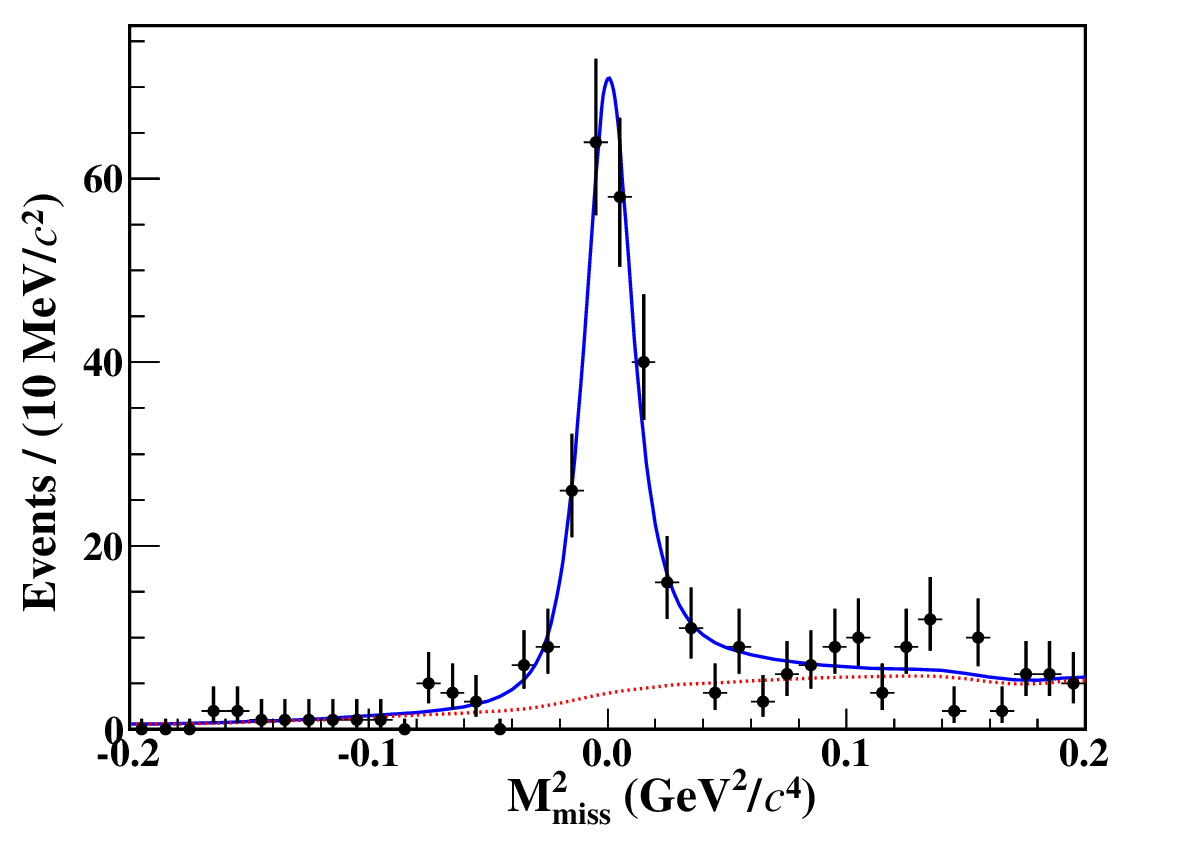}
    \end{minipage}
	\caption{\footnotesize
		Fit to the $M^{2}_{\rm miss}$ distribution of the candidates for $D_s^+\to K^0e^+\nu_e$. The points with error bars are the data summed over all CM energies; the blue solid curve is the total fit; and the red dashed curve is the fitted background shape.
	}
	\label{fig:dtfit}
\end{figure}

The $M^{2}_{\rm miss}$ distribution of the accepted candidates for $D_s^+\to K^0e^+\nu_e$ in data summed over all CM energies
is shown in Fig.~\ref{fig:dtfit}.
The signal yield of $D_s^+\to K^0e^+\nu_e$($N_{\rm DT}$) is derived from an unbinned maximum likelihood fit to this distribution.
In this fit, the signal is described by a simulated signal shape convolved with a Gaussian function with free parameters
to compensate for the resolution difference between data and MC simulation. The background is described by a simulated shape derived from the inclusive MC sample.
From this fit, the signal yield is $225.3\pm17.3$ where the uncertainty is statistical only. The corresponding DT efficiencies $\epsilon_{\rm DT}^{i}$ of various ST are summarized in the fourth column of Table~\ref{tab:bf}.

\section{Branching fraction}

\begin{figure}[htbp]
	\centering
	\includegraphics[width=0.5\textwidth]{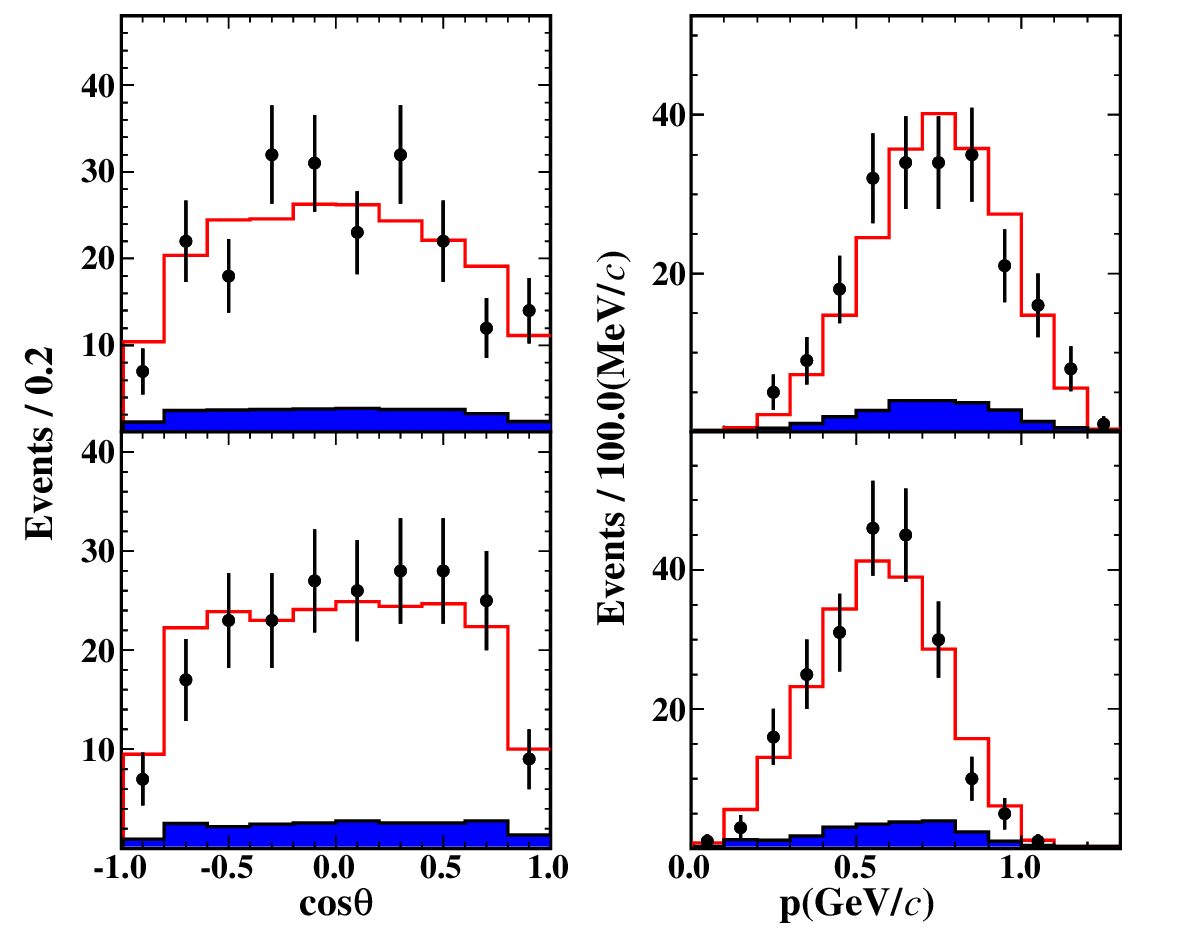}
	\caption{\footnotesize Comparison of $\cos\theta$ and momenta for the $K^0$ (top) and candidates for $D^+_s\to K^0e^+\nu_e$~(bottom) using all CM energies between 4.128 and 4.226 GeV.  The points with error bars are data; the blue filled histograms are the simulated background; and the red line histograms are the inclusive MC samples. These events have been required to satisfy $|M^2_{\rm miss}|<0.03$ GeV$^2/c^4$.
	}
	\label{fig:fit}
\end{figure}

The detection efficiency $\varepsilon_{\rm SL}$, which does not include the
branching fraction of $K^0\to K_S^0 \to \pi^+\pi^-$, is estimated to be $(27.88\pm0.21)\%$ for
$D_s^+\to K^0e^+\nu_e$.
Figure~\ref{fig:fit} shows good consistency in the $\cos\theta$ and momenta distributions for the $K^0$ and $D^+_s\to K^0e^+\nu_e$ candidates between data and the inclusive MC sample.
The branching fraction  of $D_s^+\to K^0e^+\nu_e$ is determined by Eq.~(\ref{eq1}) to be
\begin{equation}
\mathcal B({D^+_s\to K^0 e^+\nu_e})=(2.98\pm0.23\pm0.12)\times10^{-3},
\nonumber
\end{equation}
where the first uncertainty is statistical and the second is systematic, which is discussed below and summarized in Table~\ref{tab:sys_tot}.

Our measurement is performed using the DT technique~\cite{DTmethod}, and most systematic uncertainties related to the ST selection criteria therefore cancel. The systematic uncertainty of the ST $D^-_s$ yields is evaluated to be 1.0\% by using alternative signal and background shapes in the fits to the $M_{\rm tag}$ spectra.
The systematic uncertainty for the $e^{\pm}$ tracking and PID efficiency is 1.0\% each, and is studied using a control sample of $e^+e^-\to \gamma e^+e^-$~\cite{BESIII:Dsetapenu}. The systematic uncertainty in the $K_{S}^{0}$ reconstruction efficiency
is estimated with the control samples $J/\psi\to K^{*}(892)^{\mp}K^{\pm}$ and $J/\psi\to \phi K_S^{0}K^{\pm}\pi^{\mp}$~\cite{sysks} and is determined to be 1.5\% per $K^0_S$.
The systematic uncertainty of the transition $\gamma$ reconstruction~\cite{BESIII:2011ysp}, which is weighted by the branching fraction of $D_s^{*+} \to \gamma D_s^{+}$, is assigned to be 1.0\%.
The systematic uncertainties due to the requirements of $E_{\rm{extra} \gamma }^{\rm{max}}$ and $\chi^{2}_{\rm KMFIT}$
are estimated using the control samples $D_s^+\to K_S^0K^+$ and $D_s^+\to K_S^0K^+\pi^0$.
The differences of the acceptance efficiencies between data and MC simulation,
0.8\% and 2.3\%, are assigned as individual systematic uncertainties.
The systematic uncertainty due to the different tag dependence between data and MC simulation, called the tag bias~\cite{BESIII:Dsetapenu}, is estimated to be 0.2\%.
The systematic uncertainty of the quoted branching fraction of $K^0\to K^0_S\to \pi^+\pi^-$ is 0.1\%~\cite{PDG2022}.
The systematic uncertainty arising from the fit to the $M^2_{\rm miss}$ distribution is estimated to be 0.9\%
by varying the signal and background shapes.
The uncertainty due to MC statistics is 0.2\%.
Systematic uncertainty due to the uncertainty on the form factor used in the MC simulation to determine the efficiency is estimated to be 1.4\%. This is evaluated by comparing the difference of the signal efficiencies when varying the input hadronic form factor parameter by $\pm 1\sigma$, as determined in this work listed in Table~\ref{tab:sum_formfactor}.
Adding these effects in quadrature, we obtain the total systematic uncertainty on the measurement of the branching fraction of $D^+_s\to K^0 e^+\nu_e$ to be 3.9\%. A summary of the systematic uncertainties for the branching fraction is shown in Table~\ref{tab:sys_tot}.

\begin{table}[htbp]\centering
\caption{Sources of systematic uncertainties in the branching fraction measurement.}
\small
\begin{tabular}{cc}
\hline\hline
Source & Uncertainty\,(\%)\\ \hline
Single-tag yield                                                                      &1.0\\
$e^+$ tracking                                                                        &1.0\\
$e^+$ PID                                                                             &1.0\\
$K^0_S$ reconstruction                                                     			  &1.5\\
Transition $\gamma$ reconstruction                                                    &1.0\\
$E_{\mathrm{extra}~\gamma}^{\rm max}$ and $N_{\rm extra}^{\rm charge}$ requirements   &0.8\\
$\chi^2_{\rm KMFIT}$ requirement                                                      &2.3\\
Tag bias                                                                              &0.2\\
Quoted branching fraction                                                             &0.1\\
$M^2_{\rm miss}$ fit                                                                  &0.9\\
MC statistics                                                                         &0.2\\
Hadronic form factor																  &1.4\\
\hline
Total                                                                                 &3.9\\
\hline\hline
\end{tabular}
\label{tab:sys_tot}
\end{table}

\section{Hadronic form factor}

To study the decay dynamics of the semileptonic decay $D^+_s\to K^0 e^+\nu_e$, candidates are divided according to the invariant mass squared of the $e^+\nu_e$ system ($q^2 = (E_e/c+E_\nu/c)^2+|\vec{p}_e+\vec{p}_\nu|^2$) into five intervals $(0.00, 0.35]$, $(0.35, 0.70]$, $(0.70, 1.05]$, $(1.05, 1.40]$ and $(1.40, 2.16)$~GeV$^{2}$/$c^4$.
The partial decay rate in the $i$th $q^2$ interval, $\Delta\Gamma_{\rm measured}^i$, is determined by
\begin{equation} \label{eq2}
\Delta\Gamma^{i}_{\rm measured}=N_{\mathrm{produced}}^{i}/(\tau_{D^+_s}\mathcal{B}_{K^0\rightarrow\pi^+\pi^-} N_{\mathrm{ST}}^{\rm tot}),
\end{equation}
where $N_{\mathrm{produced}}^{i}$ is the $D^+_s\to K^0 e^+\nu_e$
signal yield produced in the $i$th $q^{2}$ interval in data,
$\tau_{D^+_s}$ is the lifetime of the $D^+_s$~\cite{PDG2022}, and $N_{\mathrm{ST}}^{\rm tot}$
is the total yield of ST $D^-_s$ mesons.
The number of events produced in data is calculated as
\begin{equation}\label{eq3}
N_{\mathrm{produced}}^{i}=\sum_{j}^{N_{\mathrm{intervals}}}(\varepsilon^{-1})_{ij}N_{\mathrm{observed}}^{j},
\end{equation}
where $N_{\rm observed}^j$ is the $D^+_s\to K^0 e^+\nu_e$ signal yield observed in the $j$th $q^{2}$ interval
and $\varepsilon$ is the efficiency matrix, which also includes the effects of bin migration,
given by
\begin{equation} \label{eq4}
\varepsilon_{ij}=\sum_k
\left[(N^{ij}_{\mathrm{reconstructed}} \cdot N_{\rm ST})/(N^{j}_{\mathrm{generated}} \cdot \varepsilon_{\mathrm{ST}})\right]_k/N_{\rm ST}^{\rm tot}.
\end{equation}
Here, $N^{ij}_{\mathrm{reconstructed}}$ is the $D^+_s\to K^0 e^+\nu_e$ signal yield
generated in the $j$th $q^{2}$ interval and reconstructed in the $i$th $q^{2}$ interval,
$N^{j}_{\mathrm{generated}}$ is the total signal yield generated in the $j$th $q^{2}$ interval, and the index $k$ sums over all tag modes and energies.

\begin{figure*}[htbp]\centering
	\includegraphics[width=0.8\linewidth]{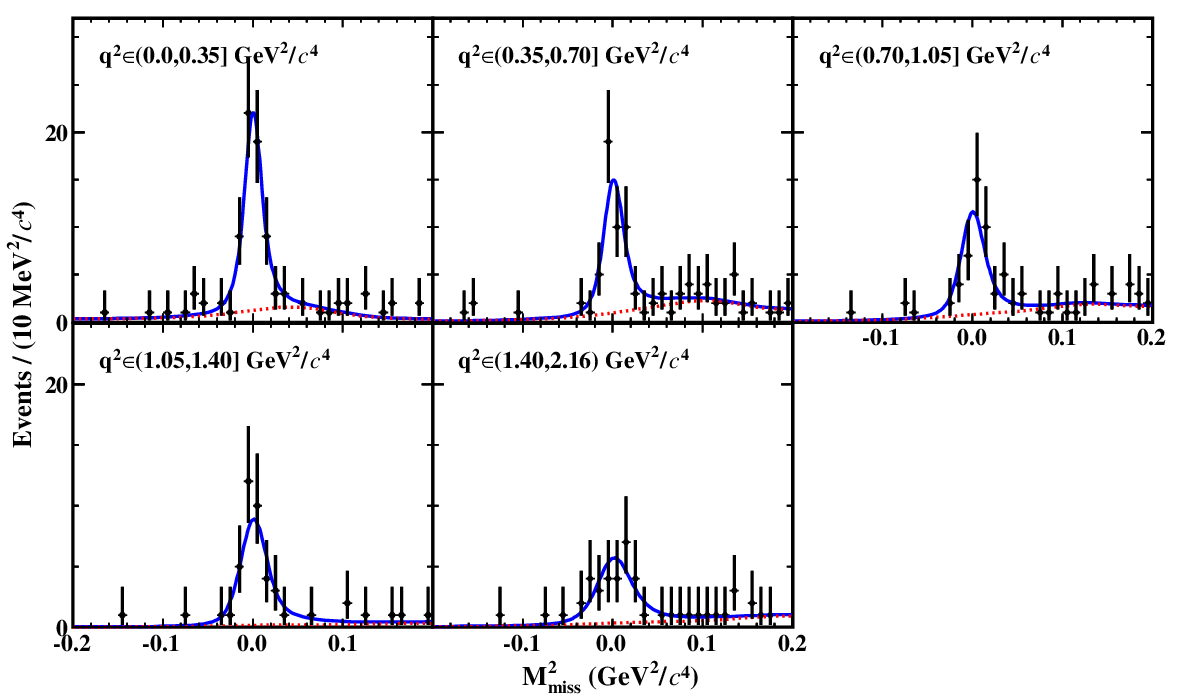}
	\caption{Fits to the $M^2_{\rm miss}$ distributions of $D^+_s\to K^0e^+\nu_e$ in various reconstructed $q^{2}$ intervals.
		The points with error bars are the data summed over all CM energies;
		the blue solid curves are the best fits;
		and the red dashed curves are the fitted background shapes.}
	\label{fig:fiteachbin}
\end{figure*}

\begin{figure*}[htbp]\centering
	\includegraphics[width=0.8\linewidth]{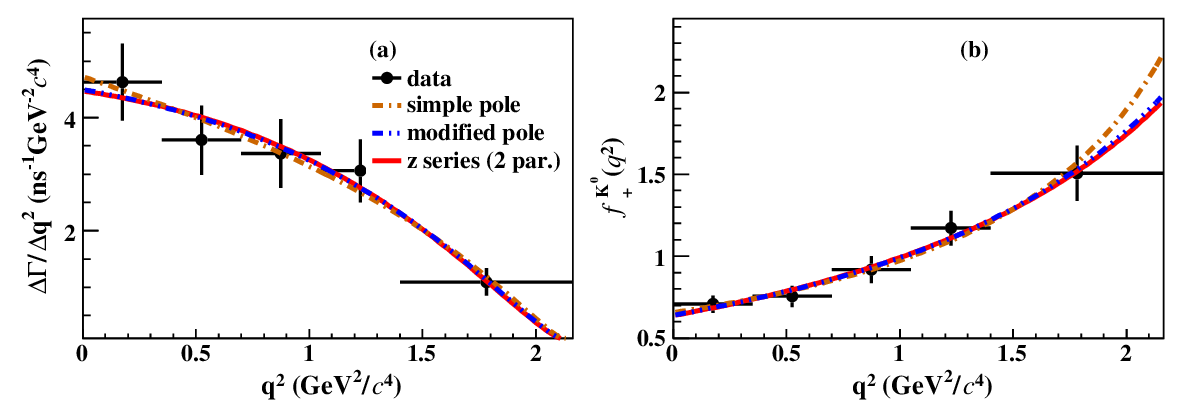}
	\caption{(a) Fit to the partial decay width of $D^+_s\to K^0e^+\nu_e$ and (b) projection to the hadronic form factor as a function of $q^2$. The points with error bars are the measured partial decay widths, where the horizonal and vertical errors represent
		the $q^2$ bin interval and the error of the corresponding partial decay width, respectively.
		The solid red curves are the best fits.}
	\label{fig:fitdecayrate}
\end{figure*}

The signal yield $N_{\mathrm{observed}}^{i}$ in each $q^2$ interval is obtained from the fit to the corresponding
$M^2_{\mathrm{miss}}$ distribution, and is shown in
Fig.~\ref{fig:fiteachbin}.
The efficiency matrix for $D_{s}^{+}\to K^{0}e^{+}\nu_{e}$ is shown in Table~\ref{table:effmatrixa}.
The values for $N^j_{\rm observed}$, $N^j_{\rm produced}$, $\Delta \Gamma_j$, and
$\frac{\Delta \Gamma_j}{\Delta q^2_j}$ are summarized in Table~\ref{table:decayrate}.

\begin{table}
	\centering
	\caption{Efficiency matrix (in \%) for $D_{s}^{+}\to K^{0}e^{+}\nu_{e}$. The efficiencies do not include the branching fractions of the decays of the daughter particles.}
	\label{table:effmatrixa}
	\begin{tabular}{c|ccccc}
		\hline\hline
		$(i,j)$ $q^2$ interval &1&2&3&4&5\\
		\hline
		1   &26.80  &0.83   &0.00   &0.00   &0.00   \\
		2   &0.93   &27.62  &0.74   &0.02   &0.00   \\
		3   &0.00   &0.91   &27.09  &0.64   &0.00   \\
		4   &0.00   &0.00   &0.83   &26.12  &0.41   \\
		5   &0.00   &0.00   &0.00   &0.70   &25.56  \\
		\hline\hline
	\end{tabular}
\end{table}

\begin{table*}[htbp]\centering
	\caption{Partial decay rates of $D_{s}^{+} \to K^{0} e^{+} \nu_{e}$ in various $q^{2}$ intervals of data, where the uncertainties are statistical only.}
	\label{table:decayrate}
	\begin{tabular}{c|ccccc}
		\hline\hline
		$q^2$ interval & (0.0,0.35]&  (0.35,0.70]&  (0.70,1.05]&  (1.05,1.40]&  (1.40,2.16]\\
		\hline
		$N^j_{\rm observed}$					& $60.7\pm8.7$ 	    & $50.8\pm8.1$ 	    & $46.1\pm7.8$ 		& $40.0\pm6.7$ 		& $30.2\pm6.5$		\\
		$N^i_{\rm produced}$					& $221.1\pm32.5$ 	& $172.2\pm29.4$ 	& $169.9\pm28.9$ 	& $146.2\pm26.5$ 	& $114.1\pm25.4$	\\
		$\Delta \Gamma_i$ (ns$^{-1}$)			& $1.62\pm0.24$ 	& $1.26\pm0.22$ 	& $1.18\pm0.21$ 	& $1.07\pm0.19$ 	& $0.84\pm0.19$		\\
		$\frac{\Delta \Gamma_i}{\Delta q^2_i}$ (ns$^{-1}$GeV$^{-2}c^{4}$)	& $4.63\pm0.68$ 	& $3.60\pm0.62$ 	& $3.37\pm0.60$ 	& $3.06\pm0.55$ 	& $1.10\pm0.24$ 	\\
		\hline
		\hline
	\end{tabular}
\end{table*}

Using the values of $\Delta\Gamma^i_{\rm measured}$ obtained above and the theoretical parameterization of the partial decay rate $\Delta\Gamma^i_{\rm expected}$ described below, form factor parameters are extracted by a $\chi^{2}$ fit where the $\chi^{2}$ is constructed as
\begin{eqnarray}
\label{eq:chisq}
	\chi^{2}=\sum_{i,j=1}^{5}&(&\Delta\Gamma^{i}_{\mathrm{measured}}-\Delta\Gamma^{i}_{\mathrm{expected}})
	 C_{ij}^{-1} \nonumber \\
&(&\Delta\Gamma^{j}_{\mathrm{measured}}-\Delta\Gamma^{j}_{\mathrm{expected}}),
\end{eqnarray}
\hspace{-0.2cm} where $C_{ij} =
C_{ij}^{\mathrm{stat}}+C_{ij}^{\mathrm{syst}}$ is the covariance
matrix of the measured partial decay rates among $q^2$ intervals.
The differential decay rate is given by
\begin{equation}
\frac{d\Gamma(D_s^+\to K^0e^+\nu_e)}{dq^2}=\frac{G^2_F|V_{cd}|^2}{24\pi^3}p^3_{K^0}|f^{K^0}_+(q^2)|^2,
\label{eq:dgammadq2_ksev}
\end{equation}
where $p_{K^0}$ is the $K^0$ momentum in the rest frame of the $D_s^+$,
$G_F$ is the Fermi coupling constant~\cite{PDG2022},
$|V_{cd}|$ is the $c\to d$ CKM matrix element, and $f_+^{K^0}(q^2)$ is the hadronic form factor.
The scalar hadronic form factor $f_0^{K^0}(q^2)$ has been ignored because it is proportional to the positron mass squared.

The hadronic form factor, $f_+^{K^0}(q^2)$, is usually parameterized by the simple pole model, modified pole model, or
series expansion.
In the modified pole model~\cite{plb478_417},
\begin{equation}
    f^{K^0}_+(q^2) = \frac{f^{K^0}_+(0)}{(1-\frac{q^2}{M^2_{\rm pole}})(1-\alpha\frac{q^2}{M^2_{\rm pole}})},
    \label{eq:mod_pole}
\end{equation}
where $M_{\rm pole}$ is fixed to the known $D^{*+}$ mass and $\alpha$ is a free parameter. Setting $\alpha=0$~and leaving $M_{\rm pole}$ free, the simple pole model is recovered~\cite{SEM}.
Due to limited statistics, we adopt the two parameter series expansion form, which is written as
\begin{equation}
f^{K^0}_{+}(q^2)=\frac{f^{K^0}_{+}(0)P(0)\Phi(0,t_{0})}{P(q^2)\Phi(q^2,t_{0})}\cdot \frac{1+r_{1}(t_{0})z(q^2,t_{0})}{1+r_{1}(t_{0})z(0,t_{0})},
\end{equation}
where
$t_{0}=t_{+}(1-\sqrt{1-t_{-}/t_{+}})$, $t_{\pm}=(m_{D^+_s}\pm m_{K^0})^{2}$, and
the functions $P(q^2)$, $\Phi(q^2, t_0)$, and $z(q^2, t_0)$ are defined following Ref.~\cite{SEM}.

The statistical covariance matrix is constructed as
\begin{equation}
C_{ij}^{\rm stat} = (\frac{1}{\tau_{D^+_s} \cdot N_{\mathrm{ST}}^{\rm tot}})^{2}\sum_{n}\varepsilon_{in}^{-1}\varepsilon_{jn}^{-1}(\sigma(N_{\mathrm{obs}}^{n}))^{2},
\end{equation}
where $n$ labels the $q^2$ interval and the sum extends from 1 to 5.
The systematic covariance matrix is obtained by summing over the covariance matrix of each systematic uncertainty source. This is taken as
\begin{equation}
C_{ij}^{\mathrm{syst}}=\delta(\Delta\Gamma^{i}_{\rm measured})\delta(\Delta\Gamma^{j}_{\rm measured}),
\end{equation}
where $\delta(\Delta\Gamma^{i}_{\rm measured})$ is the systematic uncertainty of the partial decay rate in the $i$th $q^{2}$ interval.
The systematic uncertainties due to
$\tau_{D^+_s}$,
$N_{\rm ST}^{\rm tot}$,
$e^+$ tracking efficiency,
$e^+$ PID efficiency,
$K^0_S$ reconstruction efficiency,
transition $\gamma$ reconstruction,
$\chi^2_{\rm KMFIT}$ requirement,
quoted branching fraction,
and $E_{\rm extra~\gamma}^{\rm max}$ and $N_{\rm extra}^{\rm charge}$ requirements are taken to be common across all the $q^2$ intervals.
The other systematic uncertainties in the branching fraction measurement, as shown in Table~\ref{tab:sys_tot}, are determined separately in various $q^2$ intervals. The resulting statistical and systematic correlation coefficients are summarized in Tables~\ref{table:statmatrixa} and \ref{table:systmatrixa}, respectively.

\begin{table}[htbp]\centering
    \caption{The statistical correlation coefficients of the measured partial decay rate in each $q^2$ bin for $D_{s}^{+} \to K^{0} e^{+} \nu_{e}$.}
	\label{table:statmatrixa}
	\begin{tabular}{c|ccccc}
		\hline\hline
		$\epsilon_{ij}$   &1&2&3&4&5\\
		\hline
		1   &1.000   &-0.065 &0.003  &-0.000 &0.000  \\
		2   &-0.065  &1.000  &-0.060 &0.003  &-0.000 \\
		3   &0.003   &-0.060 &1.000  &-0.057 &0.002  \\
		4   &-0.000  &0.003  &-0.057 &1.000  &-0.044 \\
		5   &0.000   &-0.000 &0.020  &-0.044 &1.000  \\
		\hline
		\hline
	\end{tabular}
\end{table}

\begin{table}[htbp]\centering
	\caption{The systematic correlation coefficients of the measured partial decay rate in each $q^2$ bin for $D_{s}^{+} \to K^{0} e^{+} \nu_{e}$.}
	\label{table:systmatrixa}
	\begin{tabular}{c|ccccc}
		\hline\hline
		$\epsilon_{ij}$   &1&2&3&4&5\\
		\hline
		1   &1.000  &0.920  &0.814  &0.952  &0.590  \\
		2   &0.920  &1.000  &0.816  &0.844  &0.706  \\
		3   &0.814  &0.816  &1.000  &0.782  &0.893  \\
		4   &0.952  &0.844  &0.782  &1.000  &0.519  \\
		5   &0.590  &0.706  &0.893  &0.519  &1.000  \\
		\hline
		\hline
	\end{tabular}
\end{table}

Minimizing the $\chi^2$ constructed in Eq.~(\ref{eq:chisq}), we obtain the product, $f_+^{K^0} (0)|V_{cd}|$, and the parameters of various hadronic form factor parameterizations.
The obtained results are summarized in Table~\ref{tab:sum_formfactor} and the fit
results are shown in Fig.~\ref{fig:fitdecayrate}.
The nominal fit parameters
are taken from the fit with the combined statistical and systematic
covariance matrix, and their statistical uncertainties are taken from
the fit with the statistical covariance matrix. For each
parameter, the systematic uncertainty is obtained by calculating the
quadratic difference of uncertainties between these two fits.
Taking the CKM matrix element $|V_{cd}|=0.22486\pm0.00067$~\cite{PDG2022} as input,
we obtain $f^{K^0}_+(0)$ as summarized in the last column of Table~\ref{tab:sum_formfactor}, where the first uncertainties are statistical and the second are systematic.

\begin{table*}
\begin{center}
\caption{Hadronic form factors of $D^+_s\to K^0 e^+\nu_e$, where the first uncertainties are statistical and the second systematic.
The parameter for the two-parameter $z$ series expansion is $r_1$ and the coefficient between the two fitted parameters is given in the fourth column.
The $\chi^2/{\rm NDOF}$ is the goodness-of-fit and NDOF is the number of degrees of freedom.}
\begin{tabular}
{lccccc}
\hline\hline
Parametrization                  & $f^{K^0}_+(0)|V_{cd}|$   &  Parameter ($M^2_{\rm pole}/\alpha/r_1$)  & Coefficient & $\chi^2$/NDOF & $f^{K^0}_+(0)$             \\
\hline
Simple pole~\cite{plb478_417}    & $0.147\pm0.009\pm0.001$& $1.75\pm0.09\pm0.03$	&0.72    & $0.96/3$ & $0.654\pm0.040\pm0.004$ \\
Modified pole~\cite{plb478_417}  & $0.144\pm0.010\pm0.002$& $0.57\pm0.25\pm0.08$	&-0.81   & $0.93/3$ & $0.640\pm0.044\pm0.009$ \\
$z$ series (two par.)~\cite{SEM} & $0.143\pm0.011\pm0.003$& $-3.7\pm1.5\pm0.5$	&0.85    & $0.97/3$ & $0.636\pm0.049\pm0.013$ \\
\hline \hline
\end{tabular}
\label{tab:sum_formfactor}
\end{center}
\end{table*}

\section{Summary}

\begin{figure*}[htbp]\centering
	\begin{tikzpicture}
		\node [ above right, inner sep=0] (image) at (0,0) {\includegraphics[width=0.49\linewidth]{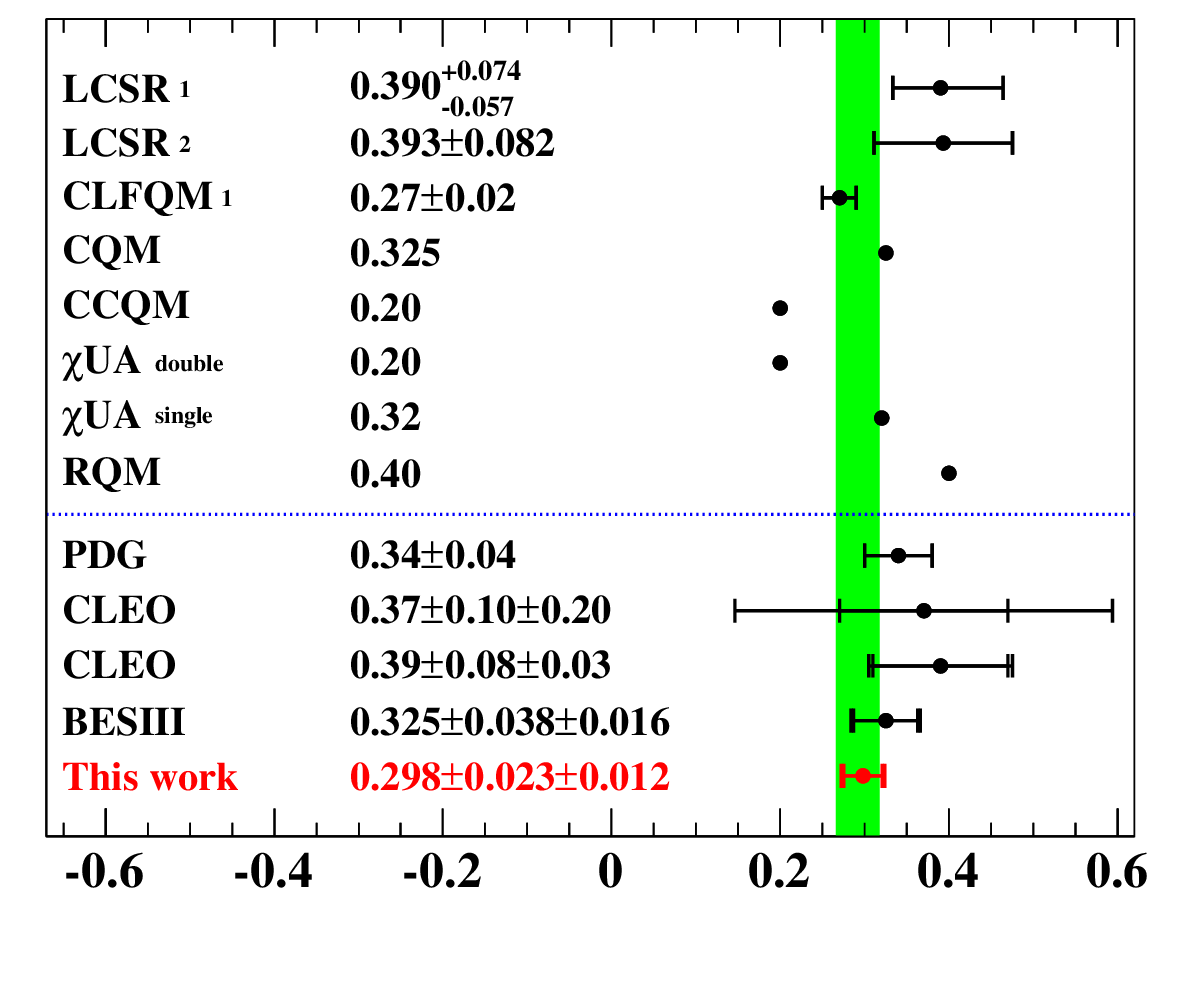}};
		\node[align=left] at (4.2,0.3) {\pmb{$\mathcal{B}(D_s^+\to K^0 e^+\nu_e)(\%)$}};
		\node [ above right, inner sep=0] (image) at (9,0) {\includegraphics[width=0.49\linewidth]{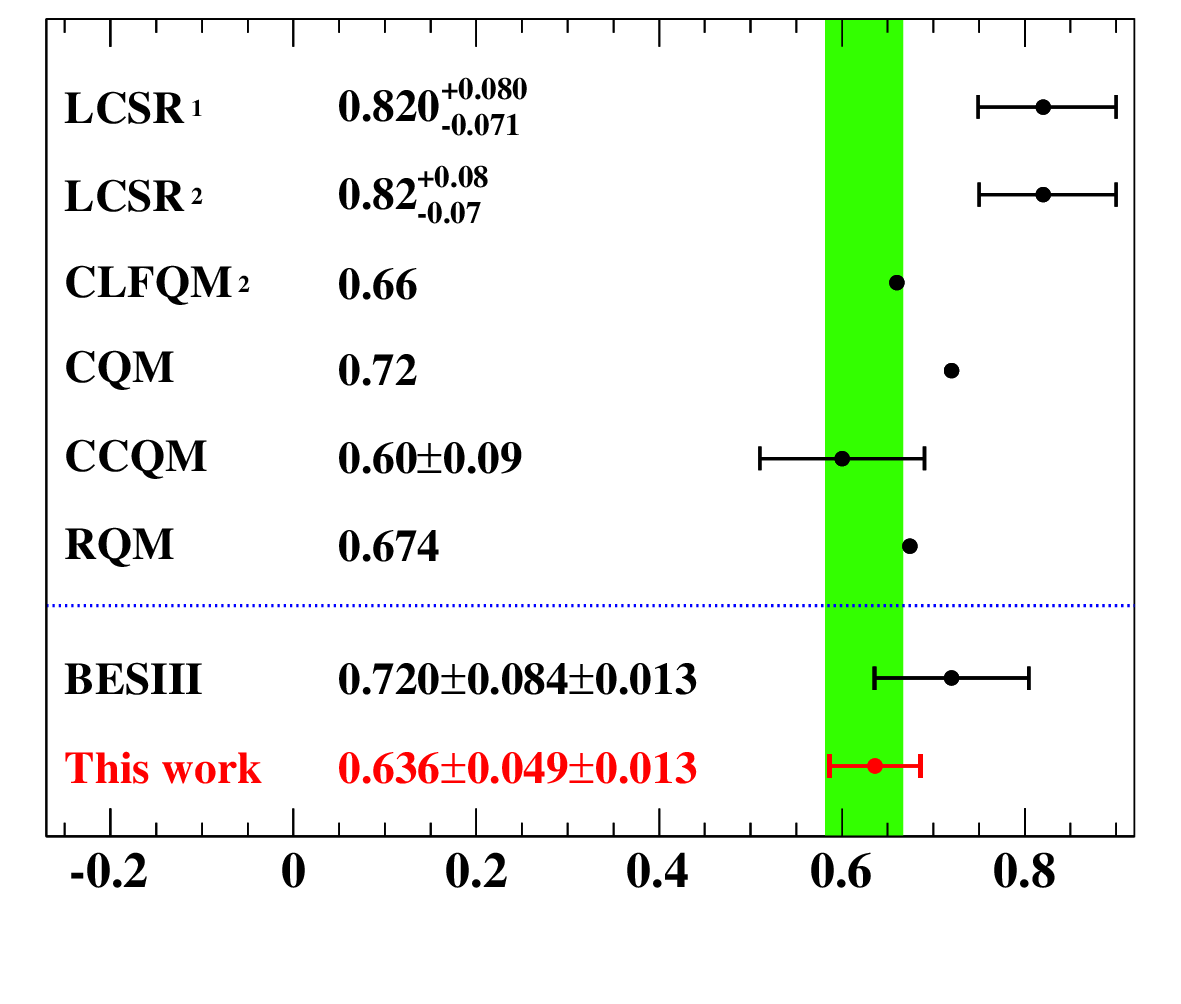}};
		\node[align=left] at (13.3,0.3) {\pmb{$f_{+}^{K^0}(0)$}};
	\end{tikzpicture}
	\caption{Comparison of the measured branching fraction of $D^+_s\to K^0 e^+\nu_e$ (left) and the hadronic form factor (right) of $D^+_s\to K^0$ with theoretical calculations and other experiments. The data in the comparison come from LCSR$_1$~\cite{Wuyl}, LCSR$_2$~\cite{ly}, CLFQM$_1$~\cite{cheng}, CLFQM$_2$~\cite{Vermayw}, CQM~\cite{Melikhov}, CCQM~\cite{soni1,soni2}, $\chi$UA$_{\rm double~pole}$~\cite{Fajfer}, $\chi$UA$_{\rm single~pole}$~\cite{Fajfer}, RQM~\cite{rqm}, PDG~\cite{PDG2022}, CLEO~\cite{cleo-K0enu}, CLEO~\cite{Hietala}, and BESIII~\cite{bes3-K0enu}.}
\label{fig:compare}
\end{figure*}

In summary, using 7.33 fb$^{-1}$ of $e^+e^-$ collision data taken between 4.128 and 4.226 GeV with the BESIII detector,
the branching fraction of $D^+_s\to K^0 e^+\nu_e$ is measured to be $(2.98\pm0.23\pm0.12)\times10^{-3}$, where the first uncertainty is statistical and the second is the systematic.
To measure the hadronic form factor at maximum recoil in $D^+_s\to K^0 e^+\nu_e$, we use the two parameter $z$ series expansion. Based on the fit to the $D^+_s\to K^0 e^+\nu_e$ partial decay rates in intervals of $q^2$, we measure $f^{K^0}_+(0)=0.636\pm0.049\pm0.013$.
Figure~\ref{fig:compare} compares the measured branching fraction and the hadronic form factor in $D^+_s\to K^0 e^+\nu_e$ with theoretical calculations and other experiments.
The precision of the measurments are improved by factors of 1.6 and 1.7, respectively, compared to the previous BESIII result~\cite{bes3-K0enu}. The results can test various theoretical calculations.

\section{Acknowledgement}

The BESIII Collaboration thanks the staff of BEPCII and the IHEP computing center for their strong support. This work is supported in part by National Key R\&D Program of China under Contracts Nos. 2020YFA0406400, 2023YFA1606000, 2020YFA0406300; National Natural Science Foundation of China (NSFC) under Contracts Nos. 11635010, 11735014, 11935015, 11935016, 11935018, 12025502, 12035009, 12035013, 12061131003, 12192260, 12192261, 12192262, 12192263, 12192264, 12192265, 12221005, 12225509, 12235017, 12361141819; the Chinese Academy of Sciences (CAS) Large-Scale Scientific Facility Program; the CAS Center for Excellence in Particle Physics (CCEPP); Joint Large-Scale Scientific Facility Funds of the NSFC and CAS under Contract No. U1832207; 100 Talents Program of CAS; The Institute of Nuclear and Particle Physics (INPAC) and Shanghai Key Laboratory for Particle Physics and Cosmology; German Research Foundation DFG under Contracts Nos. 455635585, FOR5327, GRK 2149; Istituto Nazionale di Fisica Nucleare, Italy; Knut and Alice Wallenberg Foundation under Contracts Nos. 2021.0174, 2021.0299; Ministry of Development of Turkey under Contract No. DPT2006K-120470; National Research Foundation of Korea under Contract No. NRF-2022R1A2C1092335; National Science and Technology fund of Mongolia; National Science Research and Innovation Fund (NSRF) via the Program Management Unit for Human Resources \& Institutional Development, Research and Innovation of Thailand under Contract No. B16F640076; Polish National Science Centre under Contract No. 2019/35/O/ST2/02907; The Swedish Research Council; U. S. Department of Energy under Contract No. DE-FG02-05ER41374


\end{document}